# First Principles Calculation of Magnetocrystalline Anisotropy Energy of MnBi and MnBi$_{1-x}$Sn$_x$

A. Sakuma, Y. Manabe, and Y. Kota

Department of Applied Physics, Tohoku University, Sendai Japan

We calculated the magnetic anisotropy constant $K_u$ of MnBi using a first principles approach, and obtained a negative $K_u$ in agreeable with experimental results. Furthermore, we also found a band filling dependence indicating that a slight decrease in the valence electron number will change $K_u$ from negative to positive. When some of the Bi is replaced with Sn to decrease the valence electron number, the $K_u$ value of MnBi$_{1-x}$Sn$_x$ drastically changes to a positive value, $K_u \sim 2$ MJ/m$^3$, for $x > 0.05$.

From the view point of elemental strategies, much effort has been devoted to realize strong magnetocrystalline anisotropy in transition metal systems, because of the shortage of the rare earth elements used in permanent magnets for electric vehicles. Among transition metal systems, MnBi has been expected to be a favorable candidate as a permanent magnets[1-4] or magneto-optical medium[5-7] because of its large uniaxial magnetic anisotropy. MnBi has the particular advantage that the magnetic anisotropy constant ($K$u) increases with increasing temperature, starting from a negative value around zero temperature and becoming positive at around 100 K.[2,3] This behavior continues above room temperature and $K$u reaches about 2 MJ/m$^3$ at 300 K. In addition to the large magnetocrystalline anisotropy, recently, the high spin polarization of MnBi has attracted attention both from theoretical and experimental points of view.[8] Despite these interesting properties of MnBi, most theoretical works,[9-13] except for Refs. 8), 14), and 15), have focused on the magneto-optical properties.

In the present work, we perform a first principles calculation of the $K$u of MnBi at zero temperature and of the magnetic moment $M$ and Curie temperature $T$c. The calculated $K$u is found to be negative, in agreement with the experimental results, and the electron number dependence of $K$u indicates that a decrease in the valence electron number will change the $K$u to a positive value. Based on this observation, we examine the partial replacement of Bi atoms with Sn atoms, whose electron number is one lower, and we confirm that the $K$u value is raised to positive value at zero temperature in MnBi$_{1-x}$Sn$_x$ for x >0.05.

For the electronic structure calculation, we employ the linearized muffin-tin orbital (LMTO) method under the local spin density functional approximation. In the magnetic force theorem, the magnetic anisotropy constant is given by $K_u = (E\{\hat{a}\} - E\{\hat{c}\})/V$ where $V$ is the unit cell volume and $E\{\hat{e}\}$ is the band energy calculated with the semi-relativistic LMTO Hamiltonian including spin-orbit interaction with the magnetization pointing in the $\hat{e}$ direction. The effective exchange constants are calculated using the method developed by Leichtenstein et al.[16] From the effective exchange constants, the Curie temperatures can be estimated under the molecular field approximation. The electronic structures and $K$u values of partially substituted alloys are calculated under the coherent potential approximation (CPA) within the framework of the tight-binding LMTO method.[17,18]  The crystal structure of MnBi is a NiAs-type hexagonal structure, as shown in Fig. 1, where we also show the latest lattice constants given by Yang et al.[4]  In the electronic structure calculation, we introduce empty spheres located at the dashed circles in Fig. 1.  About $2 \times 10^5$ $k$-points are sampled in the full Brillouin zone, to obtain a sufficiently converged $K$u value.

First, we look at the fundamental properties of MnBi, such as its magnetic moment and Curie temperature.  Figure 2 shows the local density of states (DOS) of MnBi.  First, we note that the spectral shape is close to that reported by Coehoorn et al.[14]  The large spin polarization in the local Mn DOS can easily be seen.  In fact, the magnetic moments calculated on the Mn sites are up to $3.8 \mu_B$, which is also consistent with the previous experimental results.[3,19]  Because of the strong hybridization of the Bi $p$ states with the Mn $d$ states, the Bi moment is polarized in the direction opposite to that of the Mn moment, with a value of $-0.08 \mu_B$.  Figure 3 shows the effective exchange constant acting on the Mn moment, which is denoted by $J_{Mn}$ as a function of the Fermi level in the rigid band scheme.  The real Fermi level position of MnBi is located at the origin of the horizontal line, whereas the actual value of $J_{Mn}$ is 72 meV.  The Curie temperature can be estimated under the molecular field approximation from $T_C = 2 J_{Mn} / 3 k_B$, from which we have $T$c ~560 K.  This is comparable to the experimental result (700 K)[2] within a reasonable accuracy.

Next, we focus on the magnetocrystalline anisotropy energy of MnBi.  Figure 4 shows the anisotropy constant $K$u of MnBi as a function of the valence electron number in the rigid band scheme.  Note that the actual electron number of MnBi is 12 per formula unit (f.u.).  We find that the actual $K$u value of MnBi is negative, at about $-0.5$ MJ/m³.  Although the absolute value is considerably larger than the measured value of $-0.2$ MJ/m³, the magnetic easy axis is consistent with the experimental results at low temperatures.  We should emphasize here that the actual electron number of MnBi, 12 per f.u. is located just above the intersection point between the Ku curve and the horizontal axis where the sign of $K$u changes from positive to negative.  This leads us to expect that the sign of $K$u may be changeable depending on the calculation condition and method.  Actually, we have confirmed that the $K$u varies with the unit cell volume with $\partial K_u / \partial V \sim -0.3 \times 10^{-30}$ MJ/m$^6$ (= $-0.3$ (MJ/m$^3$)/ Å$^3$) from which the $K$u is found to turn positive for $V < 95.5$ Å$^3$ ($V = 97.1$ Å$^3$ in the present calculation). The method for calculating the electronic structure may also have a considerable influence on the $K$u value. However, at least the $K$u value is shown to become positive upon a slight decrease in the valence electron number.

Exploiting this fact, we proceeded to calculate the $K$u of MnBi$_{1-x}$Sn$_x$ where Bi atoms are partially replaced by Sn, whose electron number is one less than that of Bi. Here, we avoided Pb substitution because of environmental consideration. In Fig. 5, the calculated $K$u values of MnBi$_{1-x}$Sn$_x$ are plotted as a function of Sn concentration $x$. We fixed the lattice constants at those of MnBi for this substitution; we have confirmed that the lattice constants do not have a substantial influence on the results. As expected, the $K$u becomes positive upon a slight substitution of Sn for Bi. In the figure, we also show, by the dashed line, the result for MnBi based on the rigid band scheme given in Fig. 4, for comparison. Below $x = 0.01$ of Sn substitution, the slope of the $K$u values against $X$ seems to be a little bit larger than that obtained in the rigid band model. For $x > 0.01$, the behavior deviates significantly from that predicted by the rigid band model, and the $K$u value roughly remains constant at around 3 MJ/m$^3$. This may be because the atomic potentials of Bi and Sn are much different from each other, invalidating the rigid band model. From a practical viewpoint, the 10% Sn substitution may be sufficient to achieve a high magnetic anisotropy at low temperature. In Fig. 6, we show the Sn concentration dependences of the Curie temperatures and magnetic moments. The data reveal that both $J_{Mn}$ and $M$ gradually decrease with $x$. However, the variation is not so dramatic, so the substitution does not have a serious influence on the magnetic properties other than the magnetic anisotropy.

It should be noted here that experimental work[20] on the magnetic properties of MnBi$R$ ($R$ = In, Ge, and Sn) films showed that only Ge doping can maintain the hexagonal structure and induce both a Kerr rotation angle and coercivity, while the structure of a Sn doped film is a mixture of the hexagonal and cubic structures. This means that the substitution of Sn for Bi is not easy to achieve in practice, because of the different atomic radii of Bi and Sn. The experimental objective is to achieve uniform replacement of Bi atoms while maintaining the hexagonal structure. Theoretically, on the other hand, the temperature dependence of the magnetic anisotropy is the main subject to be clarified. Needless to say, this has been a common problem for decades in the study of magnetism in transition metal systems. In particular, a theoretical study of the $K$u including spin reorientation at a certain temperature (in MnBi[3]; and also MnSb[21]) is sincerely desired as a future work.

In summary, we have investigated the magnetic anisotropy constant $K$u together with the magnetic moment $M$ and the Curie temperature $T$c of MnBi using a first-principles calculation. The calculated $K$u, $M$ and $T$c are to some extent consistent with the experimental results and previous theoretical works. The dependence of $K$u on the valence electron number suggests that a slight decrease in the valence electron number will change $K$u from negative to positive. Based on this result, we have calculated the electronic structure of $MnBi_{1-x}Sn_x$ whose electron number decreases with $x$ and confirmed that the $K$u dramatically changes to a positive value of ~2 MJ/m$^3$ for $x$ >0.05, while the values of $M$ and $T$c decreases slightly.

Acknowledgments
This work was supported by JST under the Collaborative Research Based on Industrial Demand program "High Performance Magnets: Towards Innovative Development of Next Magnets."

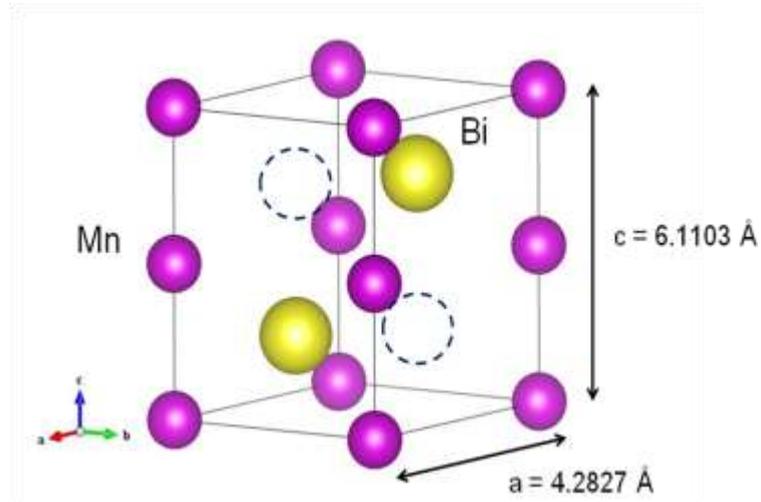

Fig. 1   Crystal structure of NiAs-type MnBi.

The dashed circles indicate the empty spheres introduced in the electronic structure calculations.

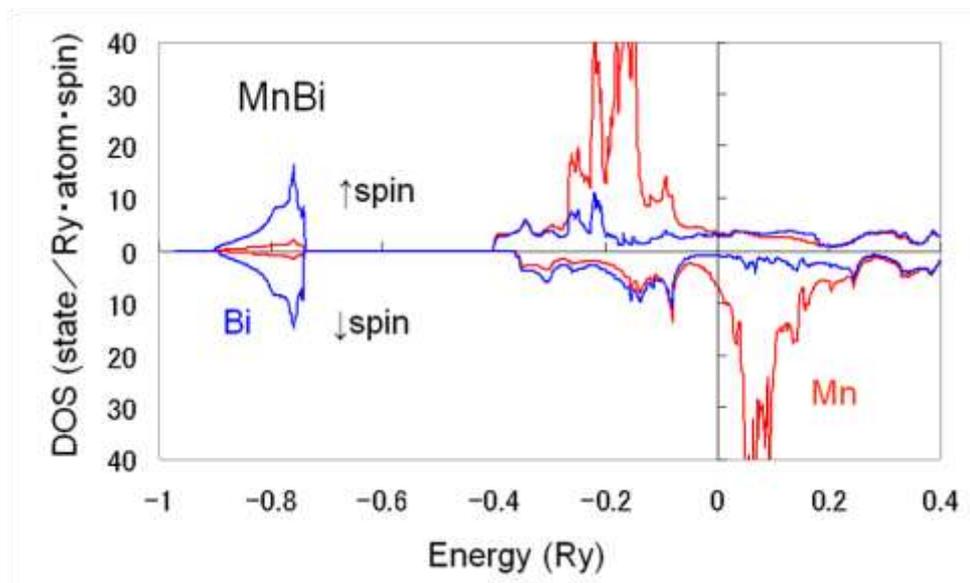

Fig. 2   Local density of states (DOS) of MnBi.   The Fermi level is located at the origin of the horizontal axis.

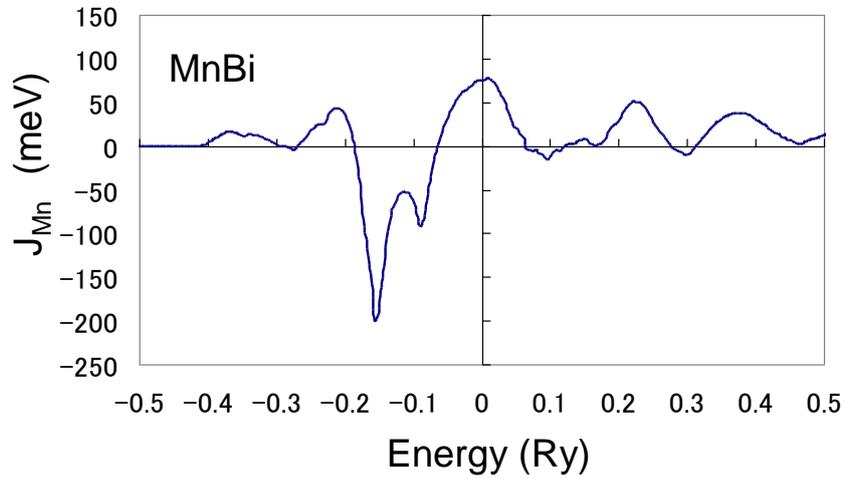

Fig. 3   Effective exchange constant $J_{Mn}$ acting on the Mn moment as a function of the Fermi level.   The actual value of $J_{Mn}$ is given at the origin of the horizontal axis.

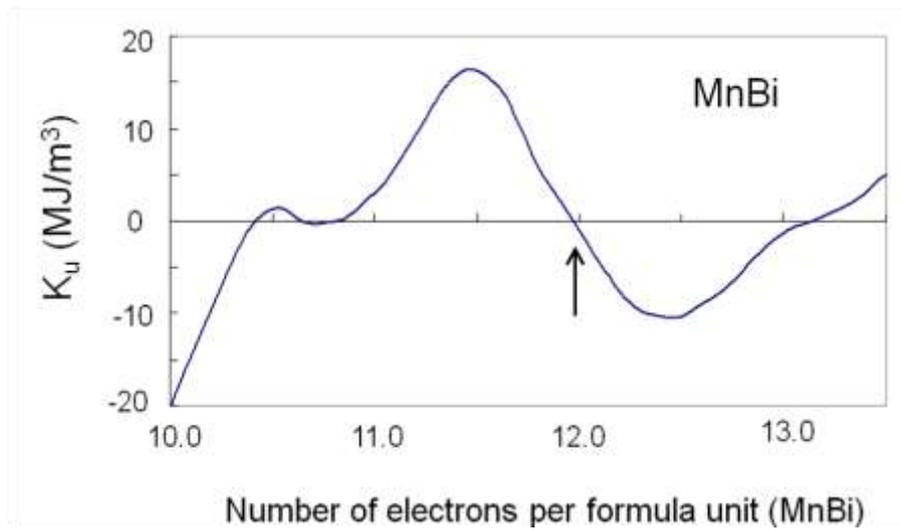

Fig. 4   Magnetic anisotropy constant $K_u$ of MnBi as a function of valence electron number in the rigid band scheme.   The actual electron number is indicated by the arrow.

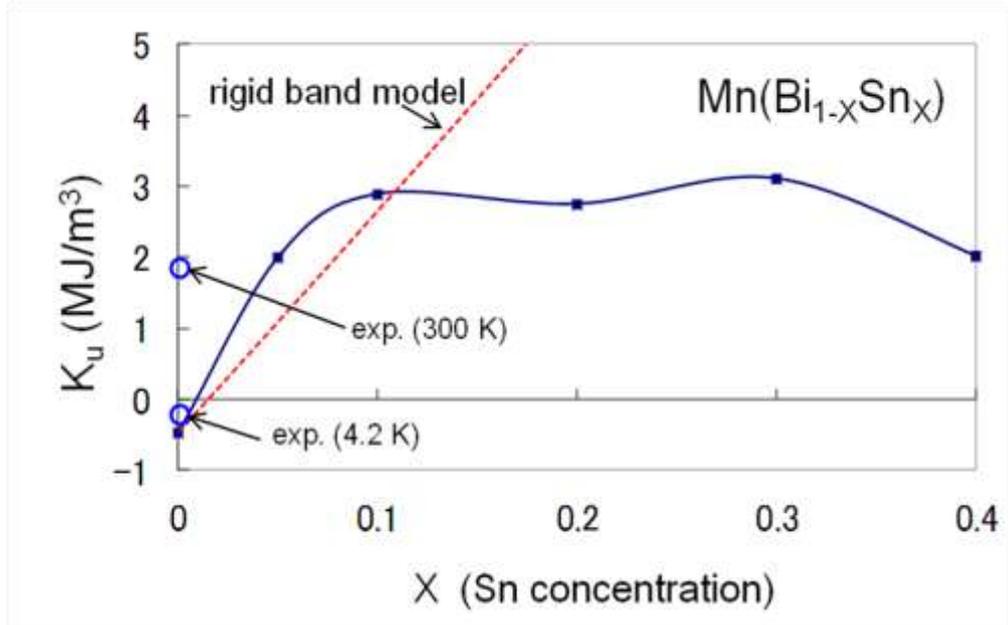

Fig. 5   $K$u of MnBi$_{1-x}$Sn$_x$ as a function of Sn concentration $x$.
The dashed line indicates the result for MnBi based on the rigid band
scheme given in Fig. 4.

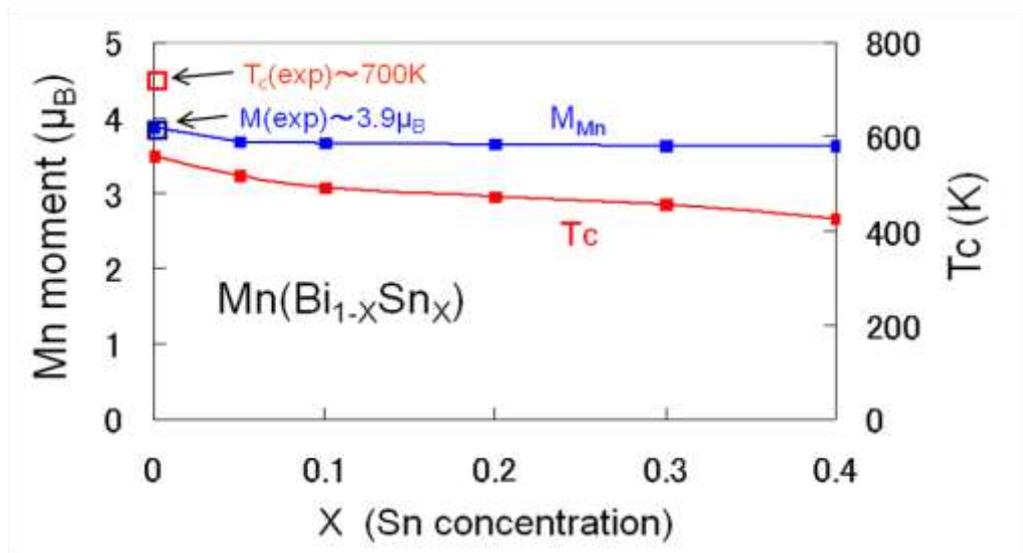

Fig. 6   $M$ and $T$c of MnBi$_{1-x}$Sn$_x$ as a function of Sn concentration $x$.